# Ternary complexes of albumin-Mn(II)-bilirubin and Electron Spin Resonance studies of gallstones

E.N. Chikvaidze[1], T.V. Gogoladze[2], I.N. Kirikashvili[1] and
G.I. Mamniashvili[3*].

[1]Javakhishvili Tbilisi State University, Georgia
[2]Tbilisi State Medical University, Georgia
[3]Andronikashvili Institute of Physics, Georgia

*Corresponding author: G.I. Mamniashvili.
E-mail address: g.mamniashvili@aiphysics.ge

ABSTRACT

The stability of albumin-bilirubin complex was investigated depending on pH of solution. It was shown that the stability of complex increases in presence of Mn(II) ions. It was also investigated the paramagnetic composition of gallstones by the electron spin resonance (ESR) method. It turned out that all investigated gallstones contain a free bilirubin radical-the stable product of its radical oxidation. Accordingly the paramagnetic composition gallstones could be divided on three main types: cholesterol, brown pigment and black pigment stones. ESR spectra of cholesterol stones is singlet with g=2.003 and splitting between components 1.0 mT. At the same time the brown gallstones, besides aforementioned signal contain the ESR spectrum which is characteristics for Mn(II) ion complexes with inorganic compounds and, finally, in the black pigment stones it was found out Fe(III) and Cu(II) complexes with organic compounds and a singlet of bilirubin free radical. It is supposed that crystallization centers of gallstones could be the polymer network of bilirubin radical polymerization in complex with different metal ions earlier discovered in gallstones.

## Introduction

Serum albumin is a main carrier of bilirubin in blood which is a toxic product of hemoglobin decomposition.

Marx [1] supposed that in the bilirubin binding with albumin the bivalent metal ions $Fe^{+2}$, $Cu^{+2}$, $Mn^{+2}$, etc. could take part as the cation-bridges.

It is known that at aggressive and chronic hepatitis, along with the increase of nonconjugated bilirubin concentration in blood, the increase of divalent manganese - Mn(II) ions concentration takes also place. Investigations carried out with radioactive manganese ions confirmed that the increase of nonconjugated bilirubin concentration correlated with Mn(II) concentration [2-4]. We have investigated the stability of albumin-Mn(II) complex depending on pH of a solution [5]. The analysis of data shows that the binding energy of Mn(II) ions with serum albumin is comparable with the energy of weak hydrogen bonding~(1-1.5) kcal/mole at pH=7.4. This indicates that the investigated complex decays easily with the varying pH of solution. The adsorption of Mn(II) ions on serum albumin depends linearly on the ionic strength of solution (Fig. 1) The increase of ionic strength of the solution decreases the pK value, but after reaching the I=0.1 value, it becomes a constant. It could be explained by a partial neutralization of charge by sodium ions in the binding centers of Mn(II) with albumin [5]. The investigation data on stability of albumin-bilirubin and albumin-Mn(II)-bilirubin complexes depending on pH of solution are shown in (Fig. 2). In the case of ternary complexes the binding constant value increases. The measurements carried out by the ESR method did not show the presence of Mn(II) ions in blood what is possibly related with their small concentration. Therefore it was decided to study the gallstones by the ESR method, as natural blood filter where it is possible the accumulation of compounds and metal ions complexes insoluble in bile.

Measurements of ESR spectra of brown pigment stones showed the presence of a high concentration of manganese ion complexes (Fig. 3 *b*) in them, although in cholesterol and black pigment stones they were not observed (Fig. 3 *a*, *c*).

During last years it has been carried out the intensive investigations of gallstones etiology but up to present time the chemical nature of the crystallization centers is not known, as well as the role of paramagnetic metal ions in their formation [6-10].

Investigations carried out by Liu X.-T. and Hu I. [11] showed that in the creation of gallstones the free bilirubin radicals should participate, which, accordingly authors' opinion, appear at interaction of metabolic free radicals, presented in living organisms, with molecules of indirect bilirubin, followed by the radical polymerization of bilirubin and formation of complexes with metal ions. On the participation of free bilirubin radicals in formation of gallstones points out also Blazovich [12]. The formation of polymeric network of bilirubin salts in gallstones was observed still in works Okubo et al. [13]. By the method of infrared spectroscopy and equilibrium swelling authors showed that polymeric network insolvable in bile is particularly strongly developed in black pigment stones.

As it was shown by ESR method, a singlet with parameters g=2.003 and ΔH=1.0 mT is observed in all investigated gallstones. This ESR signal is attributed to a free radical of the indirect bilirubin. Bilirubin is a photosensitizer. Under irradiation by visible light it takes place the photoisomerization and photooxidation of bilirubin [14-20]. As it was shown in work [5] at irradiation of its powder by blue ($\lambda_m$ =450 nm) or green ($\lambda_m$ =500 nm) light, it appears the singlet ESR signal with parameters g=2.003 and ΔH=1.0 mT (Fig. 4). The ESR signal with the same parameters appears in the bilirubin chloroform solution at aforementioned light irradiation. In (Fig. 5) it is presented a curve of accumulation of the ESR signal of free bilirubin radical at influence of light.

The ESR spectrum of brown pigment stones consists of six intensive components with splitting between them ΔH=8,7 mT and a singlet with g=2.003 and ΔH=1.0 mT (Fig. 3) characteristic for a free stable bilirubin radical. The ESR spectrum of black pigment stones besides a singlet from free bilirubin radical, consists of also from a wide structurized signal with $g_{\perp}$=2.05, $g_{//}$ =2.37 and $A_{//}$ =18.6 mT, and also an asymmetric singlet with g=4.19 (Fig. 4).

## Materials and methods

Crystal serum albumin of “Serva” firm was used without its further purification. Bilirubin of “Alfa Aesar” firm was purified on chromatographic column (L100/400 “Chemapol” Praque) [5]. The purity of preparate from free radical products of oxidation was checked by help of ESR method. In experiments it was used chromatically pure NaCl, MnCl, Tris (2-amino-2-hydroxymethyl-1,3-propanediol),pure for analysis from “Reanal” Hungary firm. The thin layer of bilirubin powder (30 mg) was irradiated by blue and green lamps “Osram” L18/67 and L18/66 (Germany) from distance 0.5 m what prevented sample’s heating. The temperature of sample was monitored by mercury thermometer with 0.1$^{o}$C accuracy. It was carried out also the irradiation of chloroform solution of bilirubin. The irradiated solution was evaporated and bilirubin powder remained after the evaporation was scraped from quartz cuvette and placed in quartz tube for ESR spectrum measurements. All bilirubin experiments were made under the red light.

The ESR spectra were taken by the 3-cm ESR-V spectrometer with high frequency magnetic field modulation. As an ESR standard it was used Mn(II) in MgO. Gallstones were cut down and placed in quartz tube in 30 mg amount for ESR spectra measurements.

The concentration of Mn(II) ion in the solution was defined by the calibrating curve.

**Discussion of results**

Accordingly the results of ESR spectra measurements the gallstones could be divided by their paramagnetic composition on cholesterol, brown pigment and black pigment stones. Cholesterol stones don't contain paramagnetic impurities and the ESR spectrum reveals only the presence of free bilirubin radical. In brown pigment stones besides this signal it is observed a complicated ESR signal consisted of six intensive components which we attribute to the hyperfine(HF) Mn(II) ions structure. As per weaker components with 2.2 mT splitting we suppose that they also belong to Mn(II) ions but only in other complexes and the actual spectrum is the superposition of spectra from different Mn(II) ion complexes. The same could be said about the ESR spectrum of black pigment stones- this spectrum is the superposition of Fe(III) ion spectra with parameter g=4.19 and the ESR spectrum characteristic for bioorganic complexes of Cu(II) with $g_{\perp}$=2.05, $g_{//}$=2.37 and $A_{//}$=18.6 mTl (Fig. 4). At the same time the ESR spectra of complexes of the same paramagnetic ions with different compounds have different hyperfine splitting and g-tensor. Therefore, for analisis of these spectra, basically for determination which paramagnetic ions participate in the formation of complexes the method of annealing was used. The samples were heated till the certain temperature in the (150$^{o}$C, 250$^{o}$C, 350$^{o}$C, 500$^{o}$C and 750$^{o}$C) areas and ESR spectra were recorded after their cooling. The received results showed that Mn(II) participated in the formation complexes in the brown pigment stones and the Cu(II) and Fe(III) ions in the black pigment stones (5). At heating of the cholesterol stones up to 500$^{o}$C the ESR spectrum is disapared. It should be noted that the ESR spectrum of brown pigment stones is absolutly identical with the ESR spectrum of chalk. At heating up to 750$^{0}$C the spectrum undergoes the same changes as one in pigment stones.

Pavel and his colleagues [21] investigated the risk of gallstones formation for the sun lovers with different types of skin. It was shown, that people with a sensitive skin (less developed pigmentary system) are exposed to a more risk of gallstones formation. The pigmentary system protects the skin from direct influence of light, but if bilirubin circulating in blood is influenced by light, the photooxidation and accumulation of the bilirubin free radicals can take place which may facilitate the formation of gallstones.

**Conclusion**

The EPR signal of stable bilirubin radical was observed by us in all types of gallstones. Even in the case when gallstones are pure holesterine stones, in their centers it is observed a brown color point and the spectrum shows a weak EPR signal of free bilirubin radical. For this reason we suppose that the crystallization center of all type gallstones is the polymeric network of free bilirubin radical in complex with bivalent metal ions and organic compounds. But different paramagnetic content of gallstones itself is defined to our opinion by the different food content of patients, specific ecology of patients environment, different illnesses experienced by patients, etc.

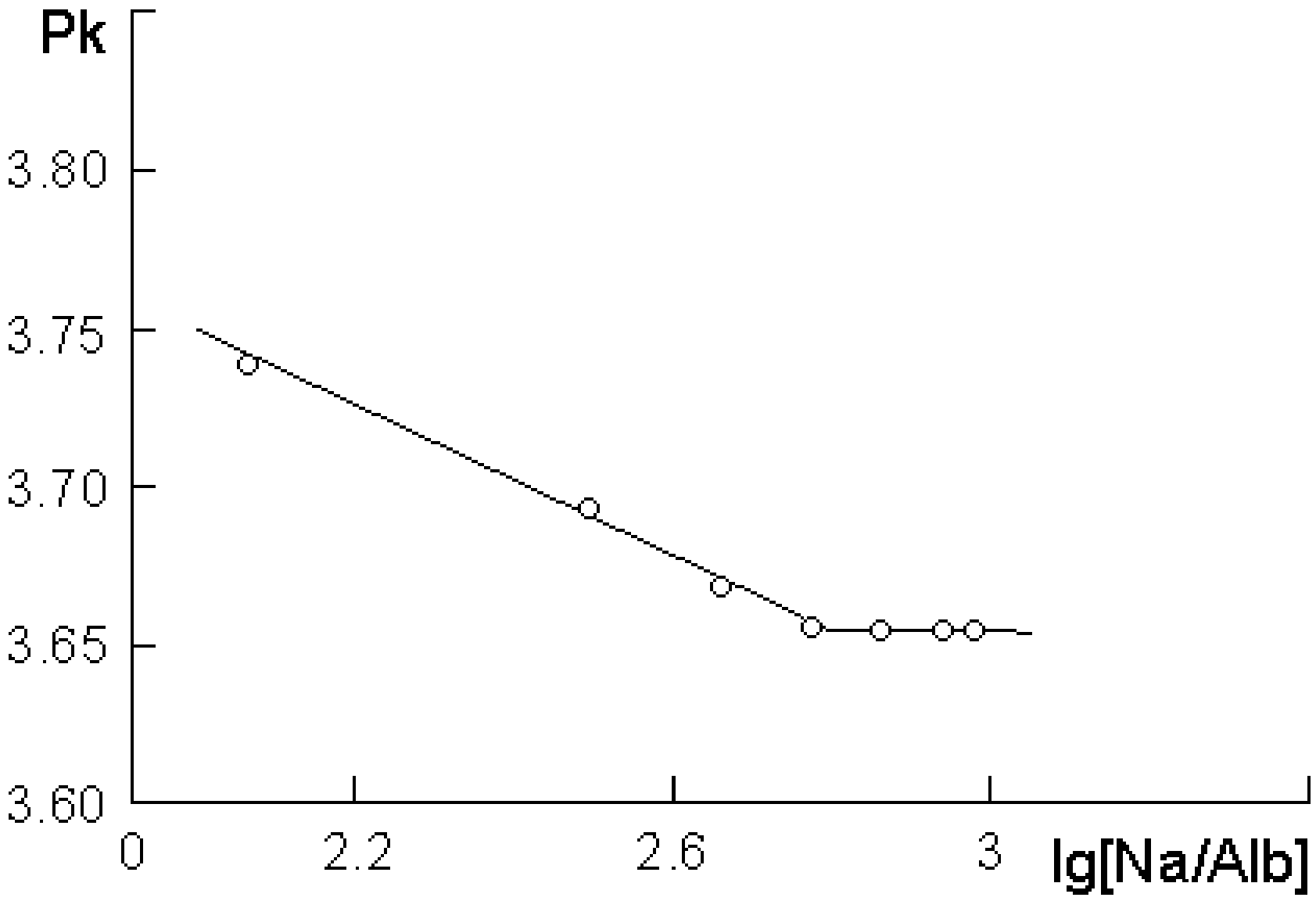

Fig. 1.

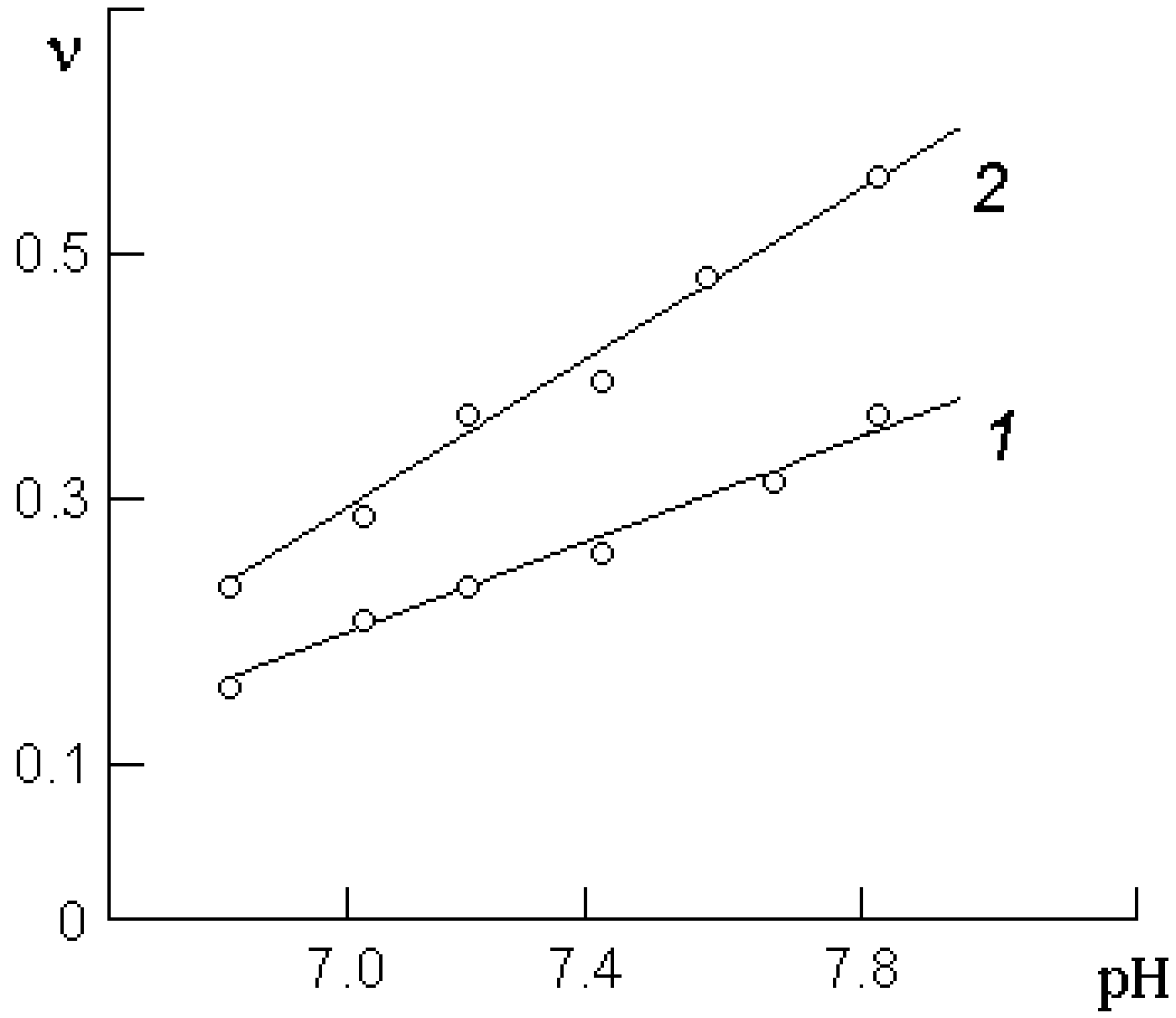

Fig. 2.

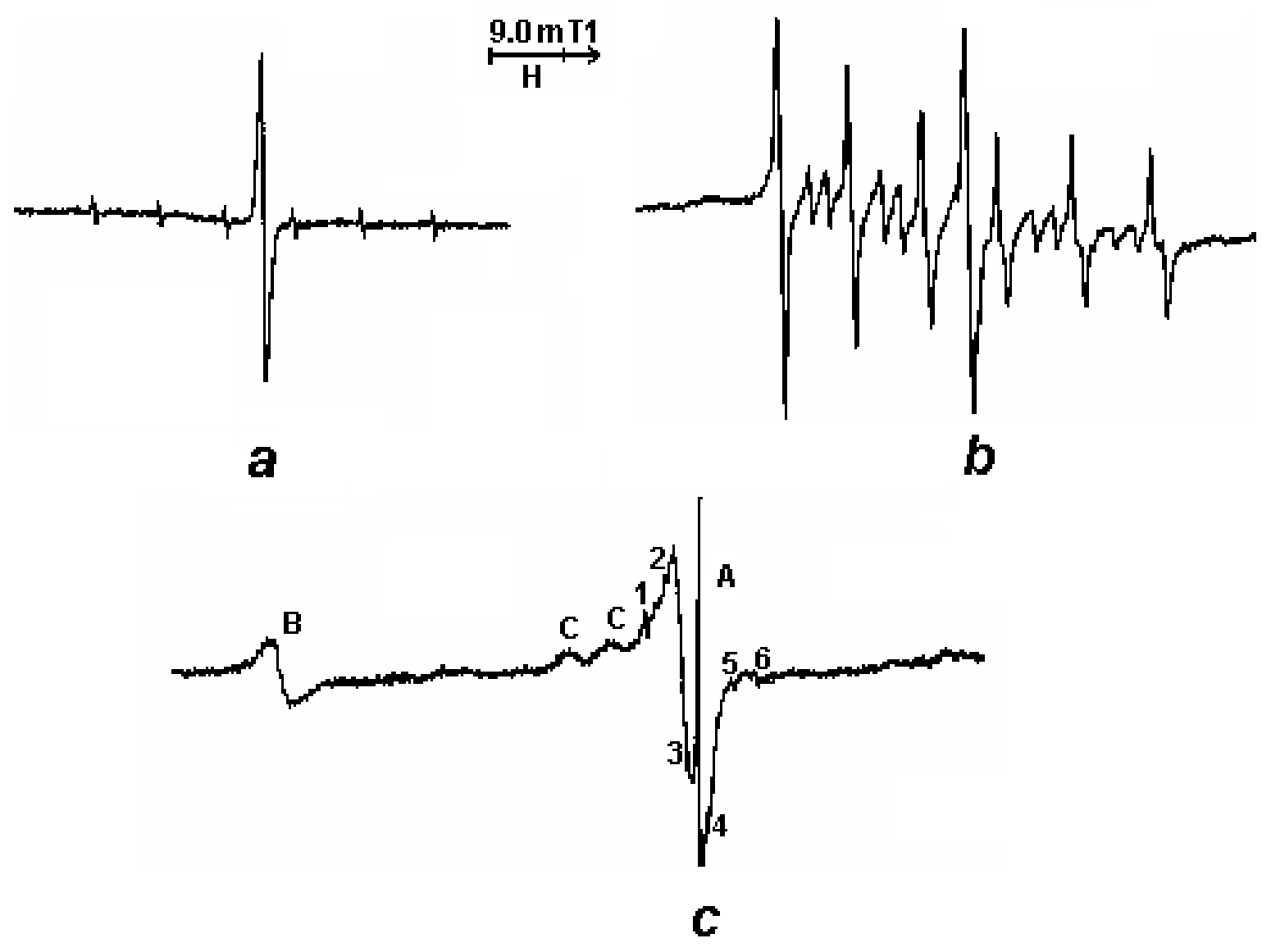

Fig. 3.

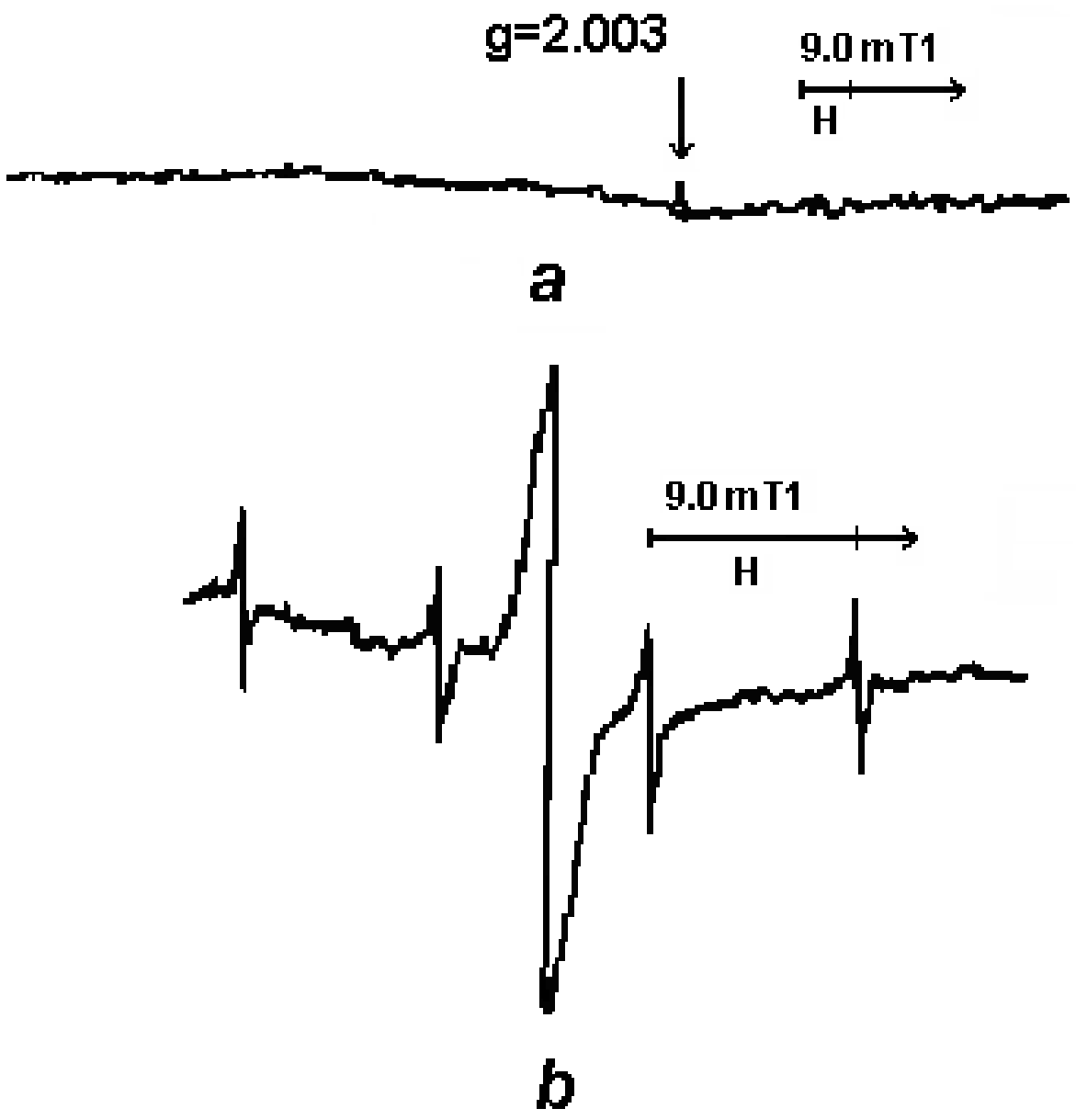

Fig. 4.

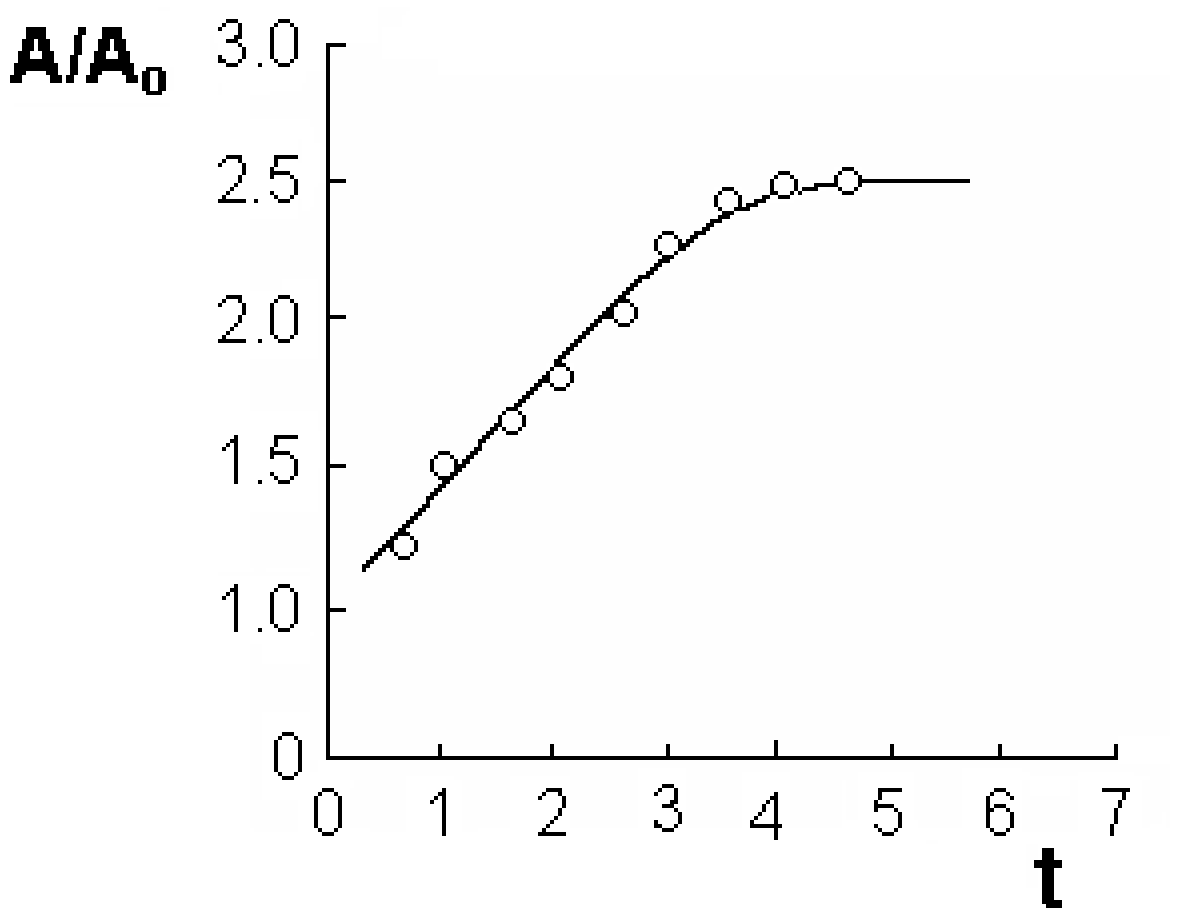
A/A0
3.0
2.5
2.0
1.5
1.0
0
0 1 2 3 4 5 6 7
t

Fig. 5

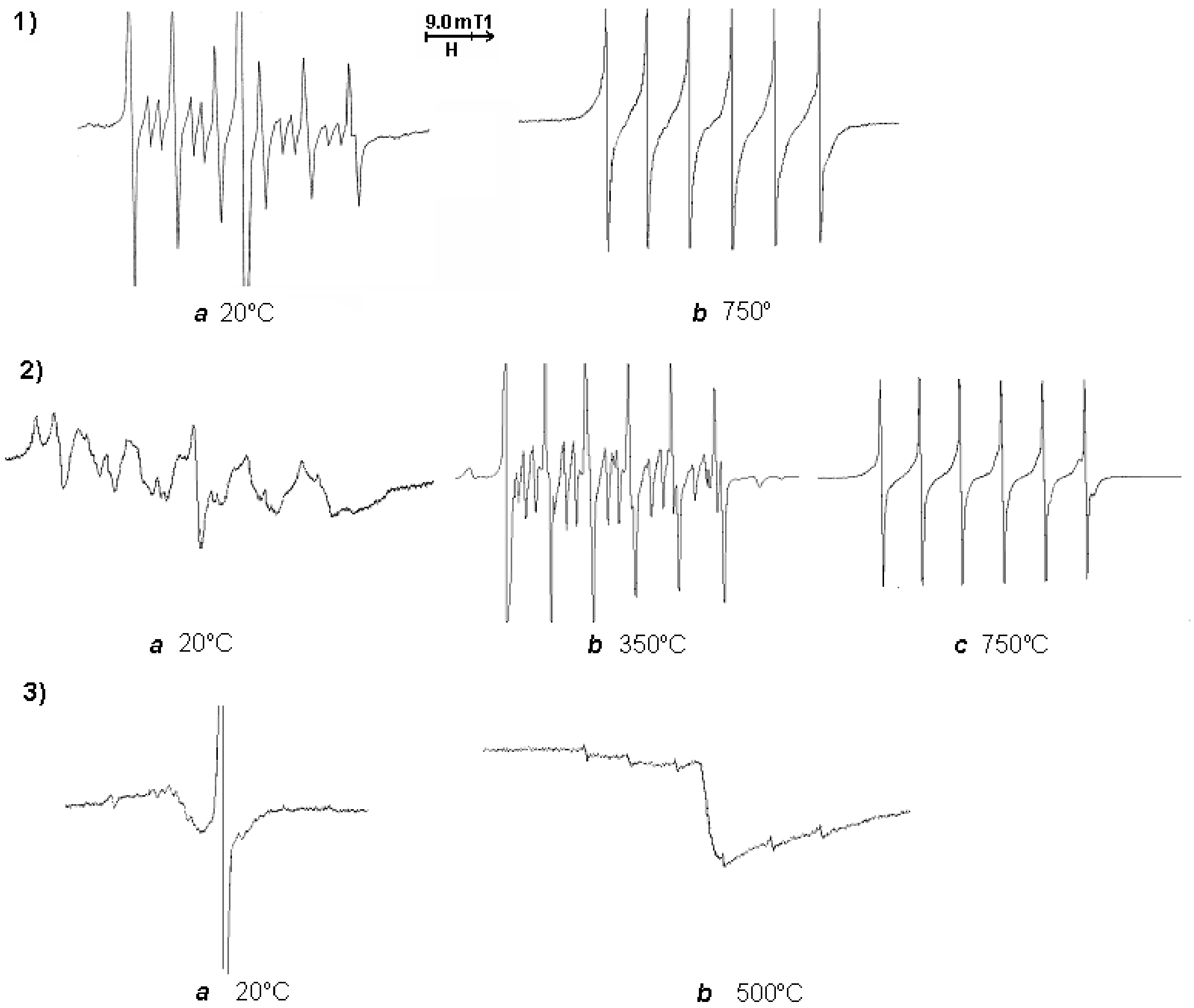

Fig. 6

# FIGURE CAPTIONS

Fig. 1. Dependence of $P_k$ of the albumin- Mn(II) complex on ionic strength of solution $[C_{Alb}]=2.0\times 10^{-4}$ M; $[C_{NaCl}]=(0.2-0.02)$ M; $[C_{Mn(II)}]=2.0\times10^{-4}$ M

Fig. 2. Plots of ν vs pH complexes: 1- albumin-bilirubin; 2 albumin-Mn(II)- bilirubin

1. The complex of albumin-bilirubin $\nu=\frac{[Br]}{[P]}B$;

2. The complex of albumin-Mn(II)- bilirubin $\nu=\frac{\lfloor Br + \mathrm{Mn(II)}\rfloor}{[P]}B$.

$[Br]_B$ -Concentration of Br bound to protein.

$[Br + \mathrm{Mn(II)}]-$ Concentration of $\lfloor Br + \mathrm{Mn(II)}\rfloor$ bound to protein.

$[P]-$ Total HSA concentration- $3.6 \times 10^{-5}$ M.

$[Br]-$ Total concentration of bilirubin -$1.8 \times 10^{-5}$ M.

[Mn (II)]- Total concentration of Mn(II) - $1.8 \times 10^{-5}$ M.

Fig. 3**.** The spectrum ESR of gallstones:

*a*) Cholesterolical stones:the singlet with g=2.003 corresponds to free a radical of bilirubin. Six HF components corresponds to Mn ( II) ions of standard;

*b*) Brown pigment stones: HFS consisting of 6-components with splitting $\Delta$ H = 8.7 mT corresponds to Mn(II) ions. The central intensive line with g=2.003 corresponds to free radical of bilirubin;

*c*) Black pigment stones: a wide line with $g_{\|} = 2.37$, $g_{\perp} = 2.05$ and HFS $A_{\|} \approx 18.6$ mT correaponds to Cu(II) complexes with organic compounds. Numbers indicate the components HFS of standard Mn(II) in MgO. The central intensive line with g= 2.003 corresponds to free radical of bilirubin.

Fig. 4**.** Spectrum ESP of pure bilirubin powder :

*a*) before irradiation;

*b*) after irradiation with blue light. The central intensive singlet belongs to a free radical of bilirubin, two HP components ( 2 and 3) belong to Mn ( II) ion of standard.

Fig. 5. The change of intensity of ESR signal of bilirubin powder irradiated by blue light . $A/A_o$ the ratio of intensivity of ESR signal of bilirubin free radical to intensity of HFS of the third component Mn ( II) ions of standard

Fig. 6. The ESR spectra of brown and black pigment stones at different temperature:

1) brown pigment stones: *a* - 20ºC, *b* - 500º C.

2) brown pigment stones: *a* - 20ºC, *b* - 350ºC, *c* - 750ºC

3) brown pigment stones: *a* - 20ºC, *b* - 500ºC.